\documentclass[apm,reprint,amsmath,amssymb]{elsarticle}

\usepackage{graphicx}
\usepackage{bm}

\begin{document}

\title{Metallic Nickel Silicides: Experiments and Theory for NiSi and
First Principles Calculations for Other Phases}

\author[um]{Ashutosh Dahal}
\author[um]{Jagath Gunasekera}
\author[nist]{Leland Harringer}
\author[um]{Deepak K. Singh}
\author[um]{David J. Singh}
\ead{singhdj@missouri.edu}

\address[um]{
Department of Physics and Astronomy, University of Missouri,
Columbia, MO 65211-7010 USA}
\address[nist]{
NIST Center for Neutron Research, Gaithersburg, MD 20899 USA}

\begin{abstract}
We report detailed experimental investigation of the transport
and magnetic properties of orthorhombic NiSi along with first principles
studies of this phase and related nickel silicides.
Neutron scattering shows no evidence for magnetism, in agreement with
first principles calculations.
Comparison of first principles results and experimental results
from our measurements and literature show a weak electron phonon
coupling. We discuss transport and other properties of NiSi
and find behavior characteristic of a weak correlated metal far from
magnetism. Trends among the nickel silicides as a function
of nickel content are discussed.
\end{abstract}

\maketitle

\section{Introduction}

Nickel silicides are important electronic materials that have been used
as contacts for field effect transistors, as interconnects and in
nanoelectronic devices.
\cite{hokelek,hu,lavoie,dong,weber,butler,gambino}
The Ni-Si binary phase diagram shows several ordered compounds all of which are
metallic except for the end point, Si.
There are both Si rich and Ni rich compounds.
The compounds include Ni$_3$Si, Ni$_{31}$Si$_{12}$, Ni$_2$Si, Ni$_3$Si$_2$,
NiSi and NiSi$_2$.
Also, although Ni is an elemental
ferromagnet, magnetic order has not been reported in the Ni-Si compounds.

The compounds in the Ni-Si phase diagram have been studied in various levels
of detail previously.
\cite{connetable,connetable-nisi,taguchi,christensen,teyssier,franciosi,butler}
This includes a study of the energetics of the
different phases by first principles calculations with the functional
known as PBESOL by Christensen and co-workers, \cite{christensen}
who obtained overbinding of the phases relative to existing thermochemical
data, \cite{oelsen1,oelsen2}
as well as investigations of the band structure, electronic density of
states and related properties. Prior work has shown all the phases to be
metallic and has investigated individual compounds to develop better
understanding of the metallic properties, 
\cite{meyer,boulet} particularly emphasizing
NiSi where quantum oscillations were measured on single crystals.
\cite{boulet}
The low resistivities that characterize the metallic state of NiSi
also transfer to nanostructures, where values as low as 10 $\mu\Omega$cm
have been obtained along with extremely high current carrying capacities.
\cite{wu}

The purpose of this paper is to report in detail the electronic and related
properties of these materials as obtained from first principles calculations,
along with an experimental investigation of the NiSi phase. This is the
most widely used phase in applications. We report structural data
consisting of fully relaxed atomic coordinates for all compounds,
energetics, that are in much better accord with experiment than prior
reports, and a detailed analysis of the electronic structures including
plasma frequencies, electronic densities of states, thermopowers and other
quantities in relation to experiment. We also report detailed
spin-polarized calculations 
including fixed spin moment calculations showing absence of magnetism.
This result is confirmed by low temperature neutron diffraction, which
proves that there is no magnetic ordering in NiSi going down to 0.48 K.
Comparison of experimental and theoretical results shows that NiSi
has weak electron phonon scattering. The results show
behavior consistent with a metallic compound far from magnetic
instabilities and in a weakly correlated regime. This provides
an explanation for the good metallic conductivity. The first principles
results suggest similar behavior for the other metallic nickel silicides,
although there is an interesting Fermi surface structure with a low
dimensional sheet in cubic Ni$_3$Si.

\section{Experimental Methods}

\begin{figure*}
\includegraphics[width=\textwidth]{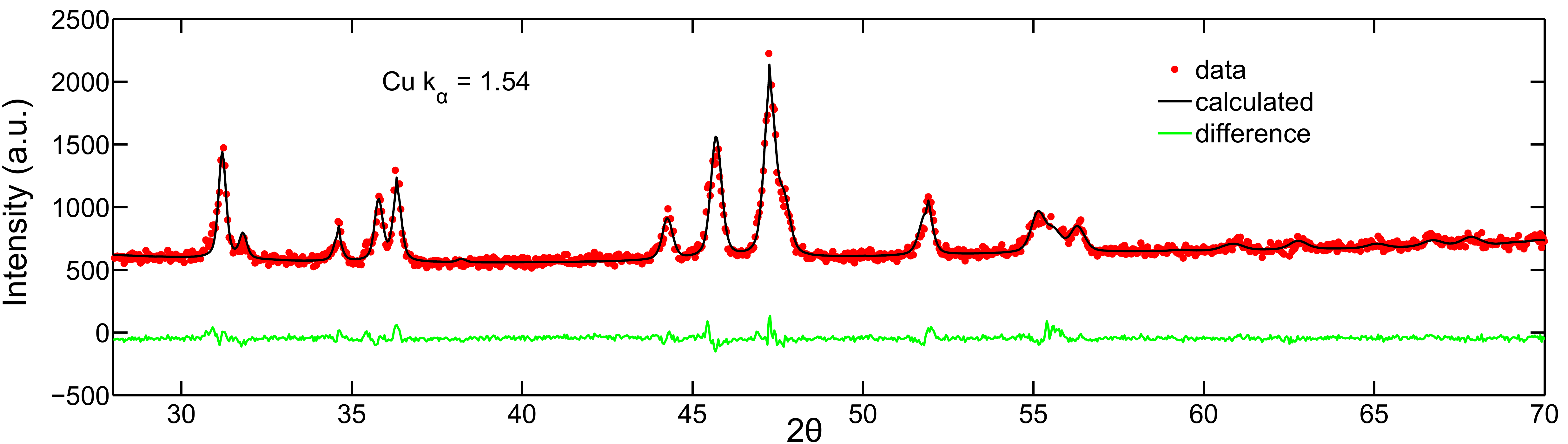}
\caption{Powder X-ray diffraction spectra of NiSi powder,
used in this study.
The powder diffraction data is refined using FullProf suite for
Rietveld analysis.
The high purity of the sample is evident from the XRD data,
where every single peak is identified to the orthorhombic structure
(crystallographic group $Pnma$) of NiSi.
}
\label{fig1}
\end{figure*}

The high purity polycrystalline samples of NiSi were synthesized by
conventional solid state reaction method using ultra-pure ingredients
(99.9\%
purity powders obtained from Alfa Aesar) of Ni and Si. Starting materials
were mixed in stoichiometric composition, pelletized and sintered for 
twenty four hours. We obtained high quality samples for sintering
temperatures of
950$^\circ$C and 970$^\circ$C.
The sample for which data is given was prepared by
(1) ramping to 970$^\circ$C over four hours, (2)
holding at 970$^\circ$C for 24 hours and
(3) cool down to room temperature over eight hours.
This sintering was done under vacuum in a sealed quartz tube.
The furnace used was a rapid temperature 1700 series front loader
manufactured by CM Inc.
The resulting samples were characterized using a Siemens D500 powder X-ray
diffractometer.
The X-ray diffraction (XRD) data was refined using the
Reitveld powder diffraction refinement.
As shown in Fig. \ref{fig1},
every single peak of the XRD pattern was identified with the
$Pnma$ orthorhombic structure of NiSi, with lattice parameters of
$a$ = 5.186 \AA, $b$ = 3.331 \AA and $c$ = 5.625 \AA.
The XRD pattern and the lattice parameters are consistent with
other reports on the synthesis of high quality samples of NiSi.
The refined structural parameters are $x_{\rm Ni}$=0.0090,
$z_{\rm Ni}$=0.1854, $x_{\rm Si}$=0.1790 and $z_{\rm Si}$=0.5874.
These are in good accord with prior data and our first
principles calculated values (see below).
The electrical properties of NiSi was characterized using the four probe
technique and utilizing a closed-cycle refrigerator cooled 9 T magnet with
measurement temperature range of 1.5-300 K.
Elastic neutron scattering measurements were performed on
the 3.6 g pristine powder sample of NiSi on the
spin-polarized triple-axis spectrometer (SPINS) at the
NIST Center for Neutron Research with fixed final neutron energy of 5 meV.
The measurements employed a flat pyrolytic graphite (PG) analyzer
with collimator sequence of 
Mono-80$^{'}$-sample-Be filter-80$^{'}$-3 blades flat
analyzer-120$^{'}$ - detector. 

\begin{figure}
\includegraphics[width=\columnwidth]{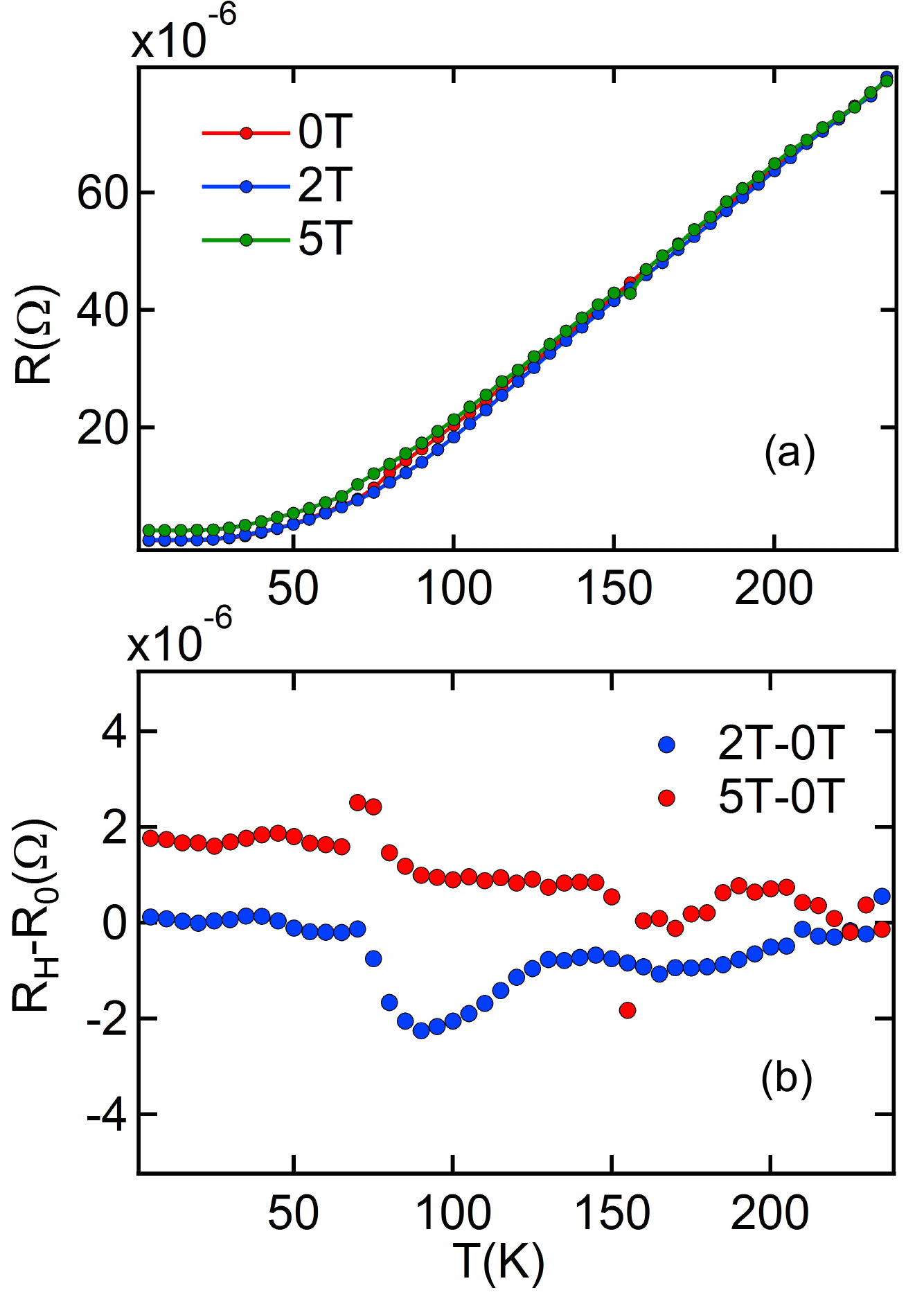}
\caption{Electrical resistance of NiSi.
(a) Electrical resistance as a function of temperature in applied field of
$H$ = 0 T, 2 T and 5 T.
(b) $\Delta$$R$ = $R$(H) - $R$(0) as a function of temperature.
}
\label{fig2}
\end{figure}

\section{Computational Methods}

The present calculations were performed within density functional theory
using the generalized gradient approximation of Perdew, Burke and Ernzerhof
(PBE-GGA), \cite{pbe} and the 
general potential linearized augmented planewave (LAPW) method \cite{singh-book}
as implemented in the WIEN2k code. \cite{wien2k}
This is an all electron method in which both core and valence states
are treated self-consistently.
LAPW sphere radii of 2.25 bohr and 1.85 bohr were used for Ni and Si,
respectively, along with well converged basis sets of the standard
LAPW plus local orbital type, with local orbitals for the Ni p semi-core state
and a planewave cutoff $k_{max}$ corresponding to $R_{min}k_{max}$=7.5,
where $R_{min}$=1.85 bohr is the Si radius (for an effective Ni value in
excess of 9). Dense Brillouin zone samplings were used and convergence with
respect to zone sampling was checked. Transport functions were calculated with
the BoltzTraP code  \cite{boltztrap} and plasma frequencies were
calculated using the optical package of WIEN2k.
In all cases we used the experimental lattice parameters and determined 
any free internal coordinates by total energy minimization.
These calculated internal coordinates are given in Table \ref{tab-struct},
along with the lattice parameters used.
For high symmetry phases, we used
Ni$_3$Si: cubic ($Pm\bar{3}m$), lattice parameter $a$=3.5098 \AA, and
NiSi$_2$: cubic ($Fm\bar{3}m$), $a$=5.406 \AA. 
\cite{weitzer,richter,richter2}

\begin{table}
\caption{Calculated internal structural parameters in lattice units
with the literature experimental lattice parameters for
NiSi ($a$=5.1818 \AA, $b$=3.334 \AA, $c$=5.619 \AA, space group 62, $Pnma$),
Ni$_3$Si$_2$
($a$=12.229 \AA, $b$=10.805 \AA, $c$=6.924 \AA, space group 36, $Cmc21$),
Ni$_2$Si ($a$=4.992 \AA, $b$=3.741 \AA, $c$=7.061 \AA, space group 62, $Pnma$),
Ni$_{31}$Si$_{12}$
($a$=$b$=6.671 \AA, $c$=12.288 \AA, space group 150, $P321$).}
\begin{tabular}{lccc}
\hline
 & ~~~~~~$x$~~~~~~ & ~~~~~~$y$~~~~~~ & ~~~~~~$z$~~~~~~  \\
\hline
{\bf NiSi:} \\
~~Ni  $4c$ & 0.0083 & 0.2500 & 0.1880 \\
~~Si  $4c$ & 0.1789 & 0.2500 & 0.5826 \\
{\bf Ni$_3$Si$_2$:} \\
~~Ni1  $8b$ & 0.1717 & 0.1189 & 0.1193 \\
~~Ni2  $8b$ & 0.3028 & 0.2533 & 0.3086 \\
~~Ni3  $8b$ & 0.3191 & 0.0046 & 0.3068 \\
~~Ni4  $8b$ & 0.3276 & 0.3813 & 0.9978 \\
~~Ni5  $4a$ & 0.0000 & 0.9999 & 0.0574 \\
~~Ni6  $4a$ & 0.0000 & 0.2343 & 0.6199 \\
~~Ni7  $4a$ & 0.0000 & 0.2347 & 0.9964 \\
~~Ni8  $4a$ & 0.0000 & 0.3830 & 0.3095 \\
~~Si1  $4a$ & 0.0000 & 0.5924 & 0.3075 \\
~~Si2  $8b$ & 0.1517 & 0.3415 & 0.1004 \\
~~Si3  $8b$ & 0.3482 & 0.1573 & 0.0177 \\
~~Si4  $8b$ & 0.3808 & 0.4431 & 0.3081 \\
~~Si5  $4a$ & 0.0000 & 0.1562 & 0.3082 \\
{\bf Ni$_2$Si:} \\
~~Ni1  $4c$ & 0.0412 & 0.2500 & 0.7057 \\
~~Ni2  $4c$ & 0.1705 & 0.2500 & 0.0602 \\
~~Si  $4c$ & 0.2122 & 0.2500 & 0.3860 \\
{\bf Ni$_{31}$Si$_{12}$:} \\
~~Ni1  $6g$ & 0.0341 & 0.3573 & 0.1948 \\
~~Ni2  $6g$ & 0.0786 & 0.4120 & 0.4045 \\
~~Ni3  $6g$ & 0.3348 & 0.2978 & 0.3022 \\
~~Ni4  $6g$ & 0.3703 & 0.3083 & 0.0963 \\
~~Ni5  $2d$ & 1/3 & 2/3 & 0.0671 \\
~~Ni6  $2d$ & 1/3 & 2/3 & 0.5728 \\
~~Ni7  $2c$ & 0.0000 & 0.0000 & 0.4029 \\
~~Ni8  $1a$ & 0.0000 & 0.0000 & 0.0000 \\
~~Si1  $3f$ & 0.6815 & 0.0000 & 0.5000 \\
~~Si2  $2d$ & 1/3 & 2/3 & 0.7755 \\
~~Si3  $3e$ & 0.3510 & 0.0000 & 0.0000 \\
~~Si4  $2d$ & 1/3 & 2/3 & 0.2736 \\
~~Si5  $2c$ & 0.0000 & 0.0000 & 0.2073 \\
\hline
\end{tabular}
\label{tab-struct}
\end{table}

\section{Experimental Results and Discussion}

As mentioned, NiSi is a low resistance compensated metal.\cite{boulet}
Previous electrical and Hall probe measurements,
especially on the single crystal specimen, of NiSi revealed two
interesting properties:
(a) very small resistivity of $\sim$ 10 $\mu$$\Omega$ cm at room temperature,
and (b) the change in the nature of the charge carrier at $T$ $\simeq$ 40 K.
The sign of Hall coefficient, $R_H$, changes from negative, at $T$$\geq$~40 K,
to positive at $T$$\leq$~40 K.\cite{meyer}
This is not unusual in metals and also frequently
occurs in compensated semimetals.
NiSi is a compensated (even number of electrons) metal.
We performed electrical measurements in both zero and applied magnetic
field on the high quality polycrystalline sample of NiSi.
Characteristic plots of electrical resistance versus temperature in
applied fields of $H$ = 0, 2 and 5 T are shown in
Fig. \ref{fig2}a
(our sample had a complex shape and so we do not report resistivity)
It is noticed that the electrical resistance in applied fields
does not exhibit any significant departure from the zero field curve
as a function of temperature. 

Further quantitative information is obtained by plotting the difference
between the zero field and the applied field resistance data,
$\Delta$$R$ = $R$(H) - $R$(0), as a function of temperature.
As shown in Fig. \ref{fig2}b,
$\Delta$$R$ fluctuates between positive
and negative values at $H$ = 2 T.
Although the net change in the electrical resistance at this field is
small compared to the bulk value at $H$ = 0 T,
the fluctuation is more prominent between 70 K and 100 K.
Unlike the electrical behavior in 2 T field,
$\Delta$$R$ remains positive at all temperatures in 5 T field. 

\begin{figure*}
\includegraphics[width=\textwidth]{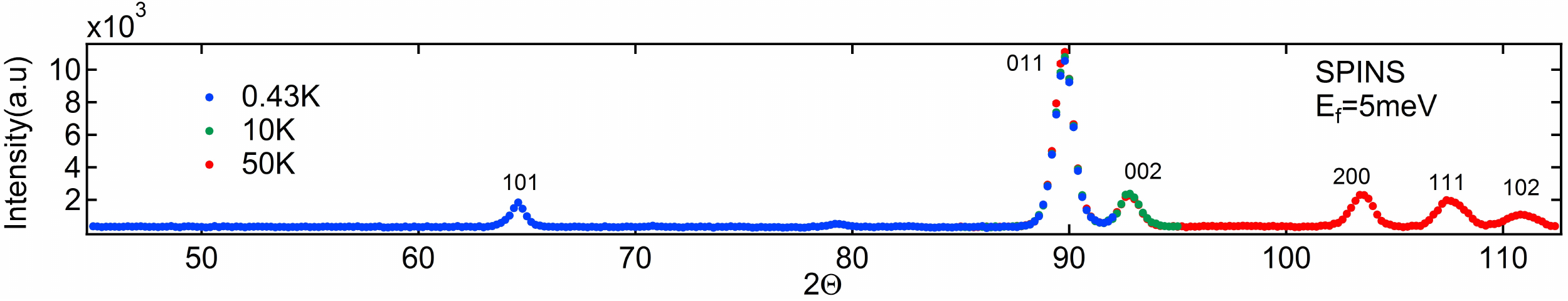}
\caption{Neutron scattering investigation of possible
magnetic order in NiSi.
Observed sharp peaks in this figure are attributed to the nuclear structure
factor. The difference between the elastic scans data at
0.48 K and 50 K is found to be statistically insignificant.
Hence, no evidence of any magnetic order is detected.
The error bars represent one standard deviation.
}
\label{fig3}
\end{figure*}

Next, we discuss the magnetic properties of NiSi.
We have performed elastic neutron scattering measurements at low temperature
up to $T$ = 0.48 K.
Elastic neutron data at few characteristic temperatures are plotted in
Fig. \ref{fig3}.
As shown in this figure,
no additional peaks are detected beyond the lattice Bragg peaks,
signifying the orthorhombic structure of the system.
Moreover, the relative intensities of the Bragg peaks
do not exhibit any temperature dependence
as the difference between the peak intensities is statistically insignificant.
This rules out any ferromagnetism in the system.
No magnetic order is detected in NiSi to
the lowest measurement temperature of $T$ = 0.48 K. 
Neutron diffraction going to very low temperature is fully
consistent with the known crystal structure and with x-ray diffraction at
high temperature. This means that there are no peaks from magnetic ordering,
which excludes any antiferromagnetic ordering. Since the material is
not ferromagnetic or ferrimagnetic this means that NiSi has no magnetic
ordering.

\section{First Principles Results and Discussion}

\begin{figure}
\includegraphics[width=\columnwidth,angle=0]{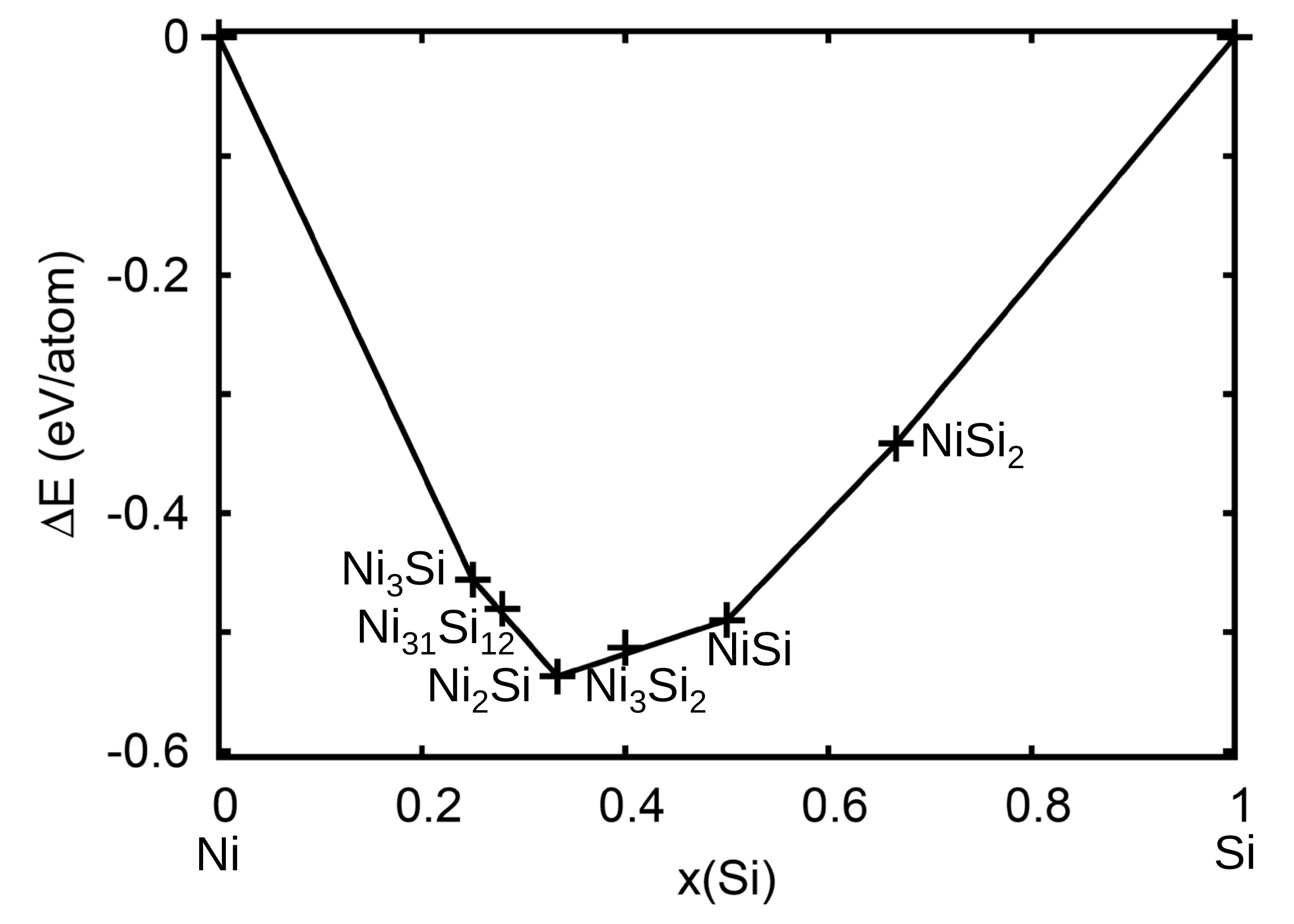}
\caption{Convex hull with calculated energies of Ni-Si compounds relative
to a mixture of the end point compounds on a
per atom basis.
The end points are ferromagnetic fcc Ni and diamond structure Si.}
\label{fig-hull}
\end{figure}

The calculated energetics are summarized in Fig. \ref{fig-hull}. All the
phases are either on the convex hull or very close to it.
All of these nickel silicide compounds are found to be metallic.
Finally the binding energies of the compounds relative to the end points
are high
reaching 0.537 eV/atom at the bottom of the convex hull. This indicates
that the compounds are very stable and that unreacted end point elements
are highly unlikely in the NiSi samples.
The calculated values deviate from reported experimental
data by $\sim$0.05 eV/atom with a systematic
overbinding of the compounds relative to that data.
\cite{oelsen1,oelsen2}
This may be within the experimental uncertainty, which has been estimated to be
0.04 eV/atom for NiSi$_2$. \cite{chandrasekharaiah}
This overbinding is much smaller than the overbinding reported in
prior calculations \cite{christensen}
that used the PBESOL functional. \cite{pbesol} PBESOL is a functional
designed to improve calculated lattice parameters at the expense of binding
energies. We verified that the functional is the origin of
the difference by repeating the calculations with the PBESOL functional for 
Ni$_3$Si and NiSi$_2$, for which we obtained an increase in the overbinding by
0.07 eV/atom and 0.11 eV/atom, respectively.
Interestingly, Ni$_{31}$Si$_{12}$ and Ni$_3$Si$_2$ are slightly above the
convex hull, but are on it to the precision expected in density functional
calculations. The phase diagram clearly shows a Ni$_3$Si$_2$ phase.
However, Ni$_{31}$Si$_{12}$ is a difficult to stabilize compound, perhaps
consistent with an energy above the convex hull.

Table \ref{tab-fermi} gives properties of the metallic state in
the various compounds.
Interestingly, the density of states at the Fermi level, $N(E_F)$,
varies by less
than a factor of two between the difference silicide compounds, but the
plasma energies, $\Omega_p=\hbar\omega_p$,
show much larger differences as well
significant but still relatively modest
anisotropy in certain compounds, specifically
Ni$_3$Si$_2$ and to a lesser extent NiSi. In metals the conductivity
$\sigma=\omega_p^2\tau$, where $\tau$ is an effective inverse
scattering rate that is both temperature and material dependent.
The inverse scattering rate in this formula
generally has weak direction dependence relative
to the plasma energy in three dimensionally bonded materials.
In this case the conductivity anisotropy has weak temperature dependence
and is given by the anisotropy of $\Omega_p^2$. In NiSi
this amounts a maximum anisotropy of $\sigma_{xx}/\sigma_{zz}$=1.47, while
in Ni$_3$Si$_2$, $\sigma_{yy}/\sigma_{xx}$=2.34.

Importantly, based on the plasma energies, NiSi, NiSi$_2$ and to a lesser
extent Ni$_2$Si are expected to be high conductivity materials, while
Ni$_3$Si, Ni$_{31}$Si$_{12}$ and Ni$_3$Si$_2$ are expected to be
lower conductivity materials. This is potentially important because
Ni$_3$Si$_2$ is the neighboring phase to NiSi in the phase diagram.
As such Ni rich NiSi material may contain Ni$_3$Si$_2$, i.e. a lower
conductivity minority phase that may have a disproportionate
effect on electrical transport. On the other hand Si rich material
would contain NiSi$_2$, which is expected to be a higher conductivity
material and have less effect on the transport.

The last column of Table \ref{tab-fermi} gives the ceramic average
values of the Seebeck coefficient at 300 K. These values were
calculated from 
$S_{av}=(\sigma_{xx}S_{yy}+\sigma_{yy}S_{zz}+\sigma_{zz}S_{zz})/
(\sigma_{xx}+\sigma_{yy}+\sigma_{zz})$,
replacing $\sigma$ by $\Omega_p^2$, i.e. using an isotropic relaxation
time.
The values of $S_{av}$ are relatively low, consistent with the 
metallic nature of these compounds. At 300 K they are weakly negative (n-type)
for all the compounds except Ni$_3$Si.

\begin{figure}
\includegraphics[width=\columnwidth,angle=0]{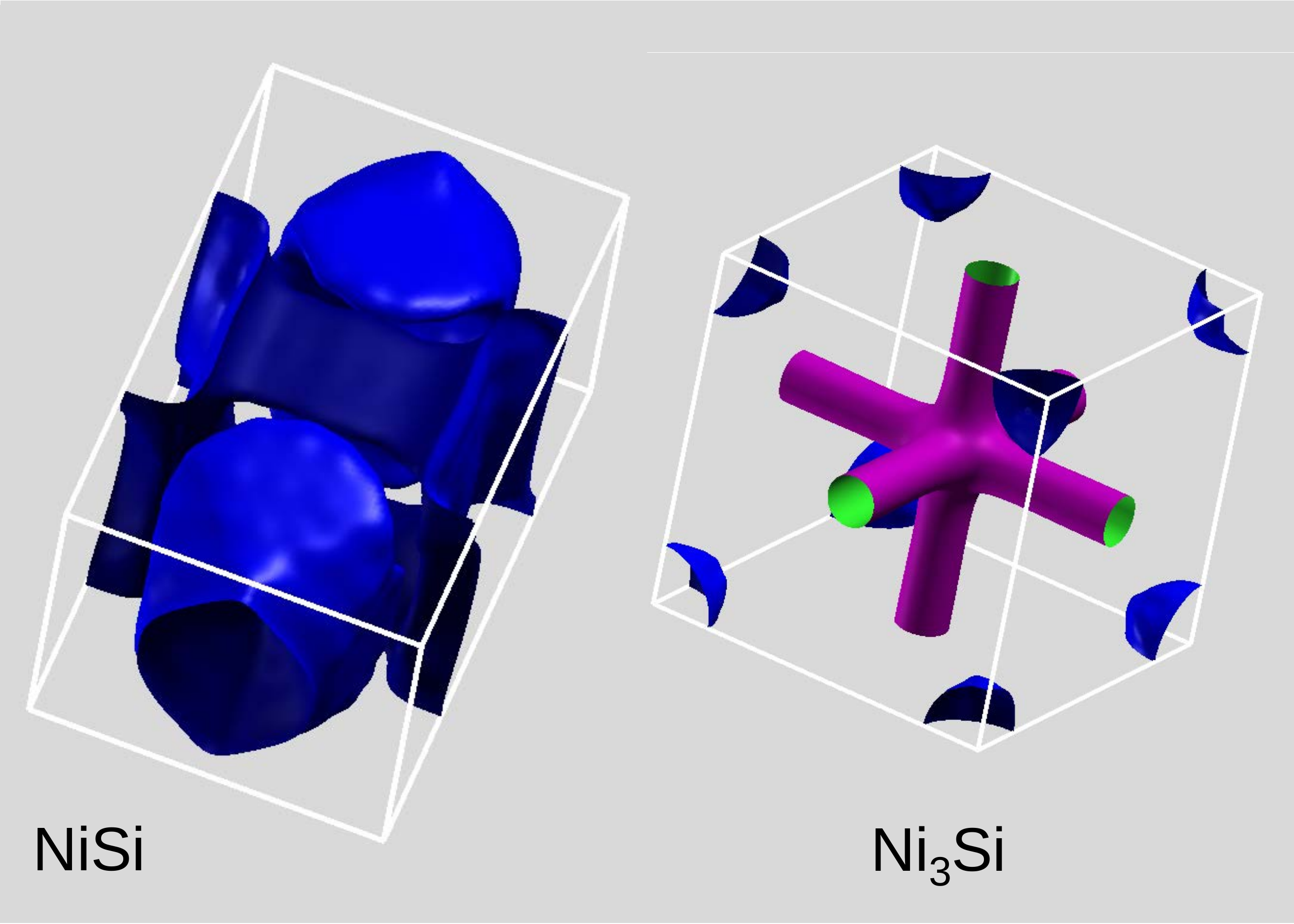}
\caption{Calculated Fermi surfaces of orthorhombic NiSi and 
cubic Ni$_3$Si. The Brillouin zones are given by the white lines.}
\label{fig-fermi}
\end{figure}

\begin{figure}
\includegraphics[width=\columnwidth,angle=0]{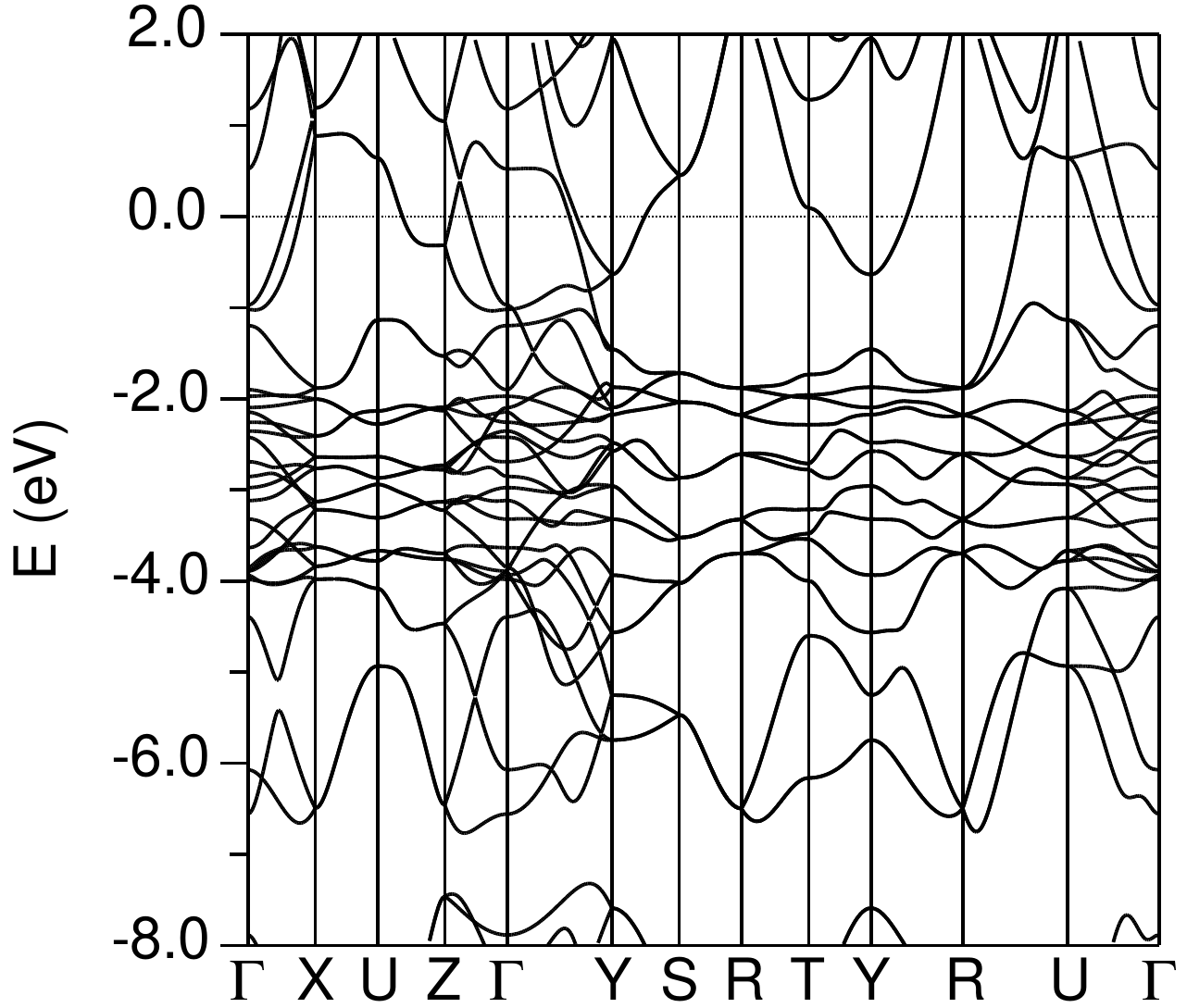}
\caption{Calculated band structure of orthorhombic NiSi.}
\label{band-nisi}
\end{figure}

\begin{figure}
\includegraphics[width=\columnwidth,angle=0]{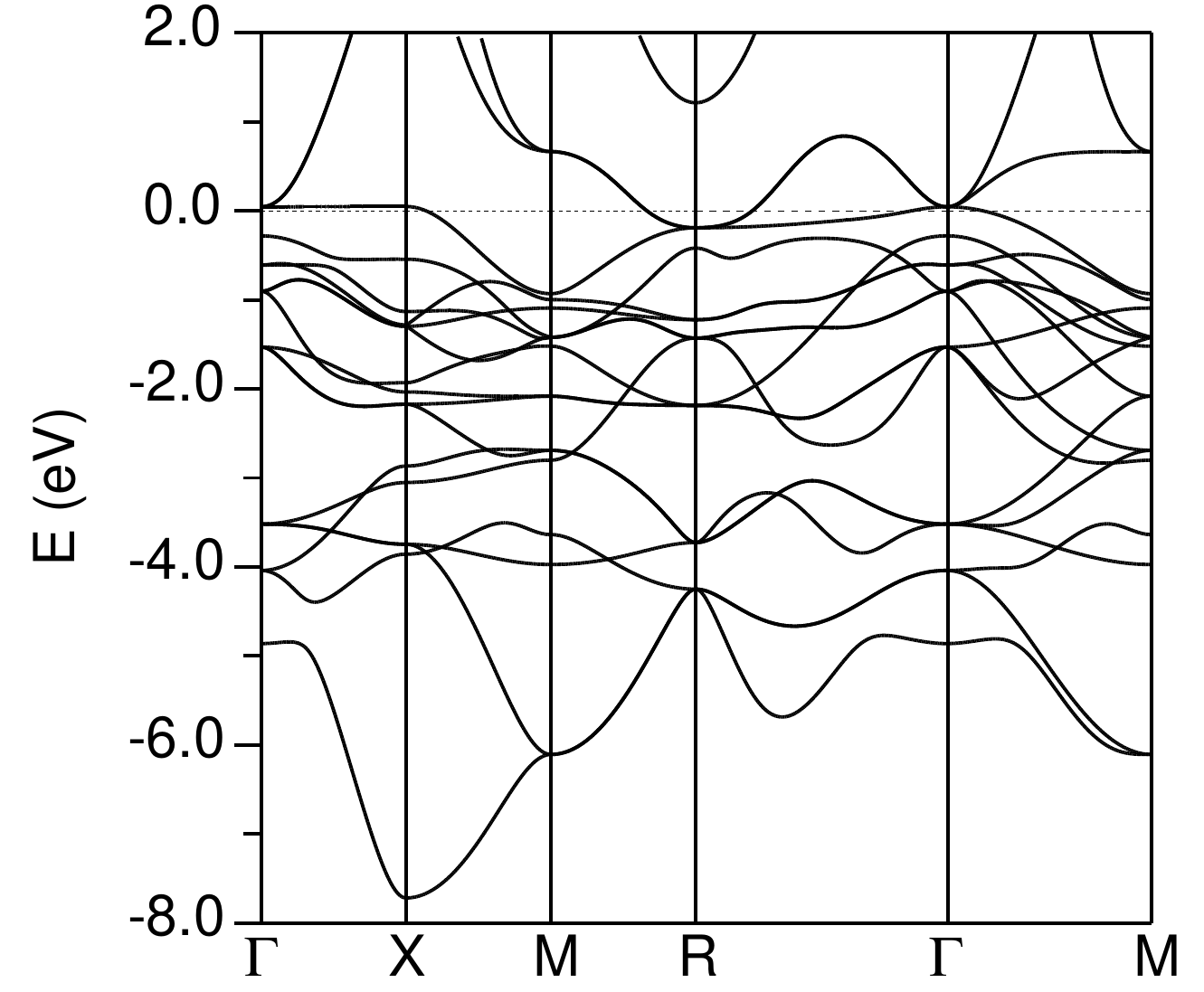}
\caption{Calculated band structure of cubic Ni$_3$Si.}
\label{band-ni3si}
\end{figure}

\begin{figure}
\includegraphics[width=0.9\columnwidth,angle=0]{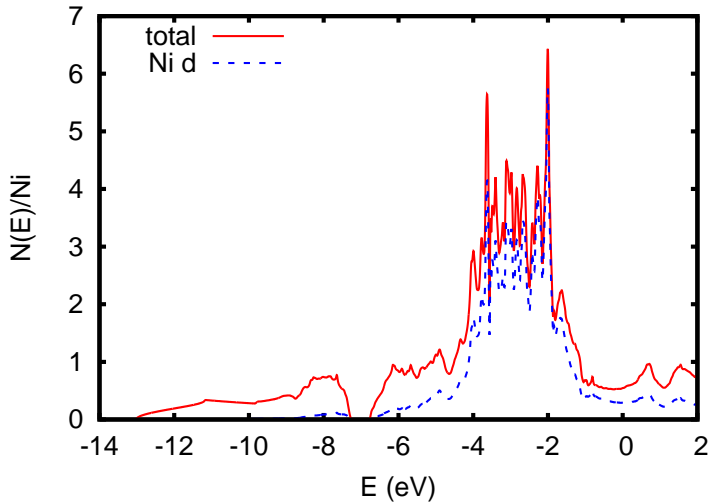}
\caption{Calculated electronic density of states and Ni $d$ contribution
on a per Ni atom basis for NiSi.}
\label{NiSi-dos}
\end{figure}

\begin{figure}
\includegraphics[width=0.9\columnwidth,angle=0]{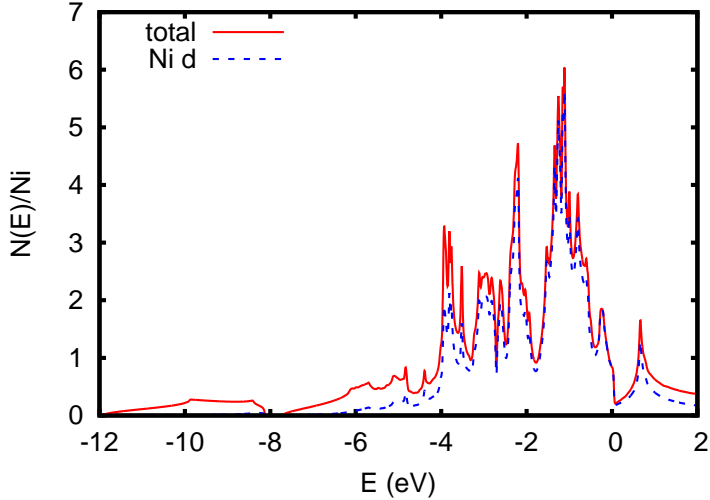}
\caption{Calculated electronic density of states and Ni $d$ contribution
on a per Ni atom basis for Ni$_3$Si.}
\label{Ni3Si-dos}
\end{figure}

The Fermi surfaces of these compounds are generally complex, large and 
for some sheets open,
reflecting the metallic nature of the materials.
Fig. \ref{fig-fermi} shows the Fermi surface of NiSi, which is the main subject
of this paper, as well as that of Ni$_3$Si, which shows an interesting
two dimensional nature for one of the sheets even though the material is cubic.
Such features have been discussed in the context of thermoelectric materials,
\cite{parker}
although Ni$_3$Si is clearly has far to low of a Seebeck coefficient to be
an interesting thermoelectric.
The corresponding
band structure of NiSi, which is similar to that reported previously
by Connetable and Thomas, \cite{connetable-nisi,connetable}
and that of Ni$_3$Si are shown in 
Figs. \ref{band-nisi} and \ref{band-ni3si}, respectively.
The open sheets of Fermi surface
in particular lead to strong anisotropy in the thermopowers,
normally the most isotropic transport property in semiconductors.
For NiSi, the calculated direction dependent thermopowers at 300 K are
$S_{xx}$=-6.0 $\mu$V/K, $S_{yy}$=-1.5 $\mu$V/K and $S_{zz}$=+0.1 $\mu$V/K,
where $x$, $y$ and $z$ are along the crystallographic $a$, $b$ and $c$
directions.
The corresponding densities of states are given in
Figs. \ref{NiSi-dos} and \ref{Ni3Si-dos}, respectively.

\begin{table}
\caption{Properties of the metallic state in nickel silicides.
The density of states at the Fermi level $N(E_F)$ is given
in $eV^{-1}$ on a
per nickel atom basis, plasma energies 
$\Omega_p$ in eV and 300K ceramic average
thermopowers, $S_{av}(300K)$ in $\mu$V/K.}
\begin{tabular}{lccccr}
\hline
 & $N(E_F)$ & $\Omega_{p,xx}$ & $\Omega_{p,yy}$ &
   $\Omega_{p,zz}$ & $S_{av}(300K)$ \\
\hline
Ni$_3$Si           & 0.83 & 1.77 & 1.77 & 1.77 &  1~~~~ \\
Ni$_{31}$Si$_{12}$ & 0.84 & 2.02 & 2.02 & 1.68 & -7~~~~ \\
Ni$_2$Si           & 0.83 & 4.25 & 4.42 & 4.00 & -15~~~~ \\
Ni$_3$Si$_2$       & 0.51 & 1.60 & 2.45 & 2.32 & -12~~~~ \\
NiSi               & 0.55 & 5.76 & 5.44 & 4.77 & -3~~~~ \\
NiSi$_2$           & 0.86 & 5.57 & 5.57 & 5.57 & -4~~~~ \\
\hline
\end{tabular}
\label{tab-fermi}
\end{table}

The density of states at the Fermi level, $N(E_F)$ for NiSi corresponds to
a bare electronic specific heat coefficient, 
$\gamma_{bare}$=1.29 mJ/(mol K$^2$).
The measured value from literature
\cite{meyer}
is $\gamma$=1.73 mJ/(mol K$^2$).
This yields an inferred enhancement, $\gamma/\gamma_{bare}=(1+\lambda)=$1.34,
i.e. a modest specific heat $\lambda$=0.34.

As mentioned, within Boltzmann theory the resistivity and plasma frequency
are related. In particular, the resistivity, $\rho^{ep}$ that comes from
electron
phonon scattering can be written

\begin{equation}
\rho^{ep}=\frac{1}{4\pi}(\Omega_p^2\tau_{ep})^{-1} ,
\end{equation}

\noindent where at high $T$

\begin{equation}
\tau_{ep}^{-1}=(2\pi\lambda_{tr}k_BT/\hbar) .
\end{equation}

\noindent Here $\lambda_{tr}$ is an electron phonon coupling constant
related to the $\lambda$ that governs electron phonon superconductivity.
\cite{allen,allen2}
Our experimental results for the resistance $R(T)$
of NiSi show a long near linear region from $\sim$75 K to at least 300 K,
consistent with prior single crystal data. 
\cite{meyer}
Meyer and co-workers reported an absolute $\rho_{xx}$=10 $\mu\Omega$cm
for current along the crystallographic $a$ axis, which is the high
conductivity axis according to the anisotropy of the calculated
plasma frequencies.
Poon and co-workers
obtained a value of $\rho$=15 $\mu\Omega$cm
for a film on polycrystalline Si.
\cite{poon}
We estimate a slope $d\rho/dT$=3.6x10$^{-10}$ $\Omega$m/K by
combining the value $\rho_{xx}$=10 $\mu\Omega$cm at 300 K with
our measured resistivity. Taking the
plasma energy, $\Omega_{p,xx}$=5.76 eV,
we obtain an estimated $\lambda_{tr}$=0.30, which is consistent with
the specific heat enhancement.
Therefore based on specific heat and resistivity in combination with the
present band structure results we infer an electron phonon coupling
constant $\lambda\sim$0.3.

This value suggests that if there are no spin-fluctuations due to
proximity to magnetism or magnetic impurities, NiSi could be a low
temperature superconductor. This is based on
normal values of the Coulomb repulsion parameter $\mu^*\sim$0.12 -- 0.15
in the Allen-Dynes equation, \cite{allen-dynes}

\begin{equation}
T_c=\frac{\omega_{eff}}{1.2}
exp \left( \frac{-1.04(1+\lambda)}{\lambda-\mu^*(1+0.62\lambda)} \right) ,
\end{equation}

\noindent where $\omega_{eff}$ is an average phonon frequency for
the spectral function. In any case, while the approximation of setting
$\lambda$ for this equation equal to $\lambda_{tr}$ or the $\lambda$
from the specific heat enhancement is clearly rough, the results suggest
that high quality NiSi samples that do not have magnetic impurities may
exhibit superconductivity. This would be
at very low temperature. With reasonable values
of $\mu^*$ and $\omega_{eff}$ one obtains $T_c$ in the 1 -- 100 mK range
due to the weak coupling.

\begin{figure}
\includegraphics[width=\columnwidth,angle=0]{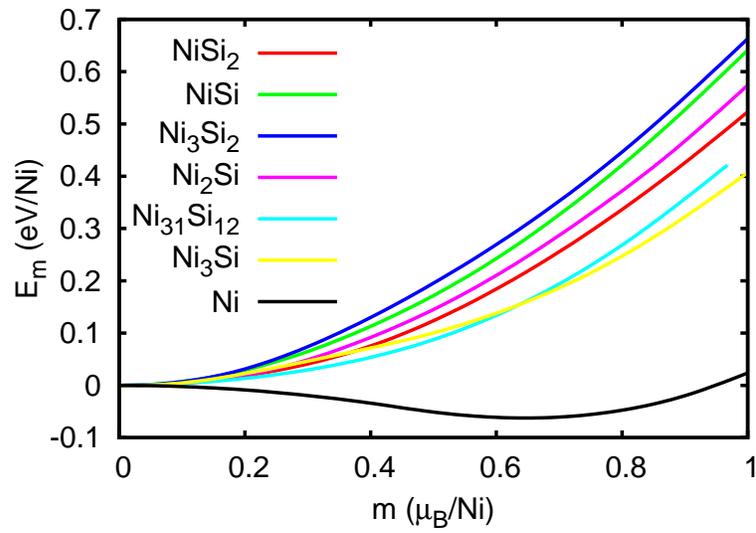}
\caption{Calculated magnetic energy as a function of constrained
spin moment from fixed spin moment calculations. The results
are given per Ni atom, both for the magnetic energy and the moment.
Elemental fcc Ni is included for comparison. Note the ferromagnetic
instability for the element in contrast to the silicides.}
\label{fig-fsm}
\end{figure}

The electronic structures show Fermi levels that are at or above
the top of the main Ni $d$ bands (see Figs. \ref{NiSi-dos}
and \ref{Ni3Si-dos}), similar to previous
density functional calculations for these materials.
\cite{connetable,connetable-nisi,taguchi,christensen,teyssier,franciosi,butler}
The results show the trend noted previously that the Ni $d$ bands move
below the Fermi level to higher binding energy as the Si content
is increased. This is important in the selection of Ni-Si contacts
for devices as it controls the level positions, and is in accord with
experimental data.
\cite{franciosi,humbert,verleysen}

As mentioned,
calculated values of $N(E_F)$ are remarkably similar between the
different compounds, varying from 0.51 eV$^{-1}$ (Ni$_3$Si$_2$) to
0.86 eV$^{-1}$, when expressed on a per Ni basis. These values are much
lower than the value for pure Ni, which is $N(E_F)$=4.9 eV$^{-1}$ for
calculations without ferromagnetism (i.e. non-spin-polarized calculations).
Within Stoner theory the proximity to itinerant ferromagnetism of a metal
is governed by a random phase approximation type expression,
$\chi=\chi_0[1-N(E_F)I]^{-1}$, where the Stoner parameter, $I$ is approximately
1 eV for late 3d transition elements, such as Ni, and $N(E_F)$ is the
density of states, expressed on a per atom, per spin basis (i.e.
half of the both spins $N(E_F)$ values given in Table \ref{tab-fermi}).
\cite{stoner,gunnarsson,janak}

In practice, $I$ is a material dependent parameter that is reduced by
hybridization between transition metal and ligand orbitals, but nonetheless
application of the Stoner formula with replacement of $N(E_F)$ by the
projection of $N(E_F)$ onto the transition metal orbitals is useful
in assessing the proximity of a given metal to a ferromagnetic 
instability. The instability occurs when $N(E_F)=I^{-1}$, i.e.
when the Ni projection of the density of states,
$N_{Ni}(E_F)$=1 eV$^{-1}$ per spin per atom.
Clearly, none of the silicide phases studied is near ferromagnetism
by this criterion, and furthermore none of them has a substantial
enhancement of their paramagnetism, governed by $[1-N(E_F)I]^{-1}$.

We did fixed spin moment calculations to verify this. The results
of such spin-polarized calculations
for all compounds except Ni$_{31}$Si$_{12}$ are shown in
Fig. \ref{fig-fsm}.
As shown, none of the compounds has a ferromagnetic tendency and
the energies as functions of the imposed spin moment are featureless,
as is often the case for simple metals that are not near magnetism.
We also performed calculations for some simple antiferromagnetic 
configurations in NiSi, but did not find any stable solutions.
These spin polarized calculations, both for initial ferromagnetic
and initial antiferromagnetic configurations always converged to
zero spin density and an energy equal to that of the non-spin-polarized
case.
The facts that the energies of non-spin-polarized and spin-polarized
calculations are the same and that the self-consistent
moment in the cell is zero confirm that there is no magnetism, consistent
with the fixed spin moment calculations.
This is
consistent with our neutron scattering results.
Spin polarized calculations for initial ferromagnetic configurations
converged to the non-spin-polarized solution for all compounds studied,
except pure Ni for which a ferromagnetic solution is found as expected.
The calculated spin moment of fcc Ni is 0.635 $\mu_B$ per atom.

Within standard Kohn-Sham density functional theory the exchange correlation
potential depends on the density and the spin-density. When the spin-density
is zero, i.e. when spin-polarized calculations converge to a non-magnetic
state, as is the case here, the exchange-correlation potential becomes
exactly the same for spin-up and spin-down. Then the
electronic density of states is exactly the same for spin up
and spin down.
Therefore the spin-up and
spin-down densities of states are identical and equal to half of the
total density of states.

\section{Summary and Conclusions}

We report experimental and theoretical study of NiSi and results of 
first principles calculations for other nickel silicides, all of which
are metallic. Neutron scattering measurements to 0.48 K confirm
that there is no magnetic ordering in NiSi.
Combination of first principles
results with experimental data show both weak electron phonon scattering
and behavior consistent with a metallic compound far from magnetic
instabilities and in a weakly correlated regime. The first principles
results suggest similar behavior for the other metallic nickel silicides,
although there is an interesting Fermi surface structure with a low
dimensional sheet in cubic Ni$_3$Si.

\section{Acknowledgements}

Theoretical work was supported by the Department of Energy, Office of Science,
Basic Energy Sciences,
through the Computational Synthesis of Materials Software Project.

% \bibliography{NiSi}

\end{document}